\def   \ni {\noindent}

\def   \ssk {\vskip  5truept}

\def   \bsk {\vskip 15truept}

\def   \newline {\hfil\break}

\documentstyle[epsfig]{article}
\begin{document}

\hsize 5truein
\vsize 8truein
\font\abstract=cmr8
\font\keywords=cmr8
\font\caption=cmr8
\font\references=cmr8
\font\text=cmr10
\font\affiliation=cmssi10
\font\author=cmss10
\font\mc=cmss8
\font\title=cmssbx10 scaled\magstep2
\font\alcit=cmti7 scaled\magstephalf
\font\alcin=cmr6 
\font\ita=cmti8
\font\mma=cmr8
\def\ref{\par\noindent\hangindent 15pt}
\null
%\vskip 3.0truecm
%\baselineskip = 12pt

% beginning of font "title"

\title{\ni CLUSTERS OF GALAXIES IN X-RAYS
}                                               

% beginning of font "author and affiliation"
\bsk \bsk
\author{\ni S.~Schindler}                                                       
\bsk
\affiliation{Astrophysics Research Institute, Liverpool John Moores
University, Twelve Quays House, Egerton Wharf, Birkenhead L14 1LD, UK 
}                                                
\bsk
\baselineskip = 12pt

% beginning of font "abstract and keywords"
\abstract{ABSTRACT \ni
X-ray observations of 
clusters at different redshifts are presented and compared.
For the nearest cluster -- the Virgo cluster -- we show a comparison of 
the galaxy distribution and the distribution of the intra-cluster gas.
Although the Virgo cluster has 
such an irregular structure, it seems that within each of the 
subclusters some kind of equilibrium has already established.
X-ray observations of distant lensing clusters reveal
very different properties. The large range of X-ray properties without
any obvious correlation between them shows that cluster evolution is
complex, i.e. not every cluster is evolving in the same way. 
A comparison of nearby clusters and distant clusters shows
that there is no clear trend with time in any of the X-ray properties,
which favours a low $\Omega$ universe.

}                                                    
\bsk
\baselineskip = 12pt

\bsk
\baselineskip = 12pt

% beginning of font "text"

\text{\ni 1. WHY ARE X-RAY OBSERVATIONS OF GALAXY CLUSTERS INTERESTING?
\ssk
\ni     
Galaxies are not the only observable component in clusters, but there
is hot gas between the galaxies. The gas is an important component
as its mass amounts to about 5 times the mass in the galaxies. The temperature
of the gas is typically between 2 and 10 keV so that it is emitting
thermal bremsstrahlung in X-rays.
From X-ray observations of clusters one can learn a variety of things
out of which I can mention only a few here.

\noindent
-- Cluster detection is much easier in X-rays than in optical, because
   the X-ray emission is proportional to the square of the gas
   density, while in optical one sees only the galaxy
   density. Therefore, the detection is much less affected by
   projection effects, as one does not know a priori which of the
   galaxies observed in the direction of the cluster are really
   cluster galaxies. Moreover, the gas fills the potential well
   of the cluster throughout, so that one can see the structure of the
   cluster at first glance (see Fig.~1).

%--------------------------  figure 1
\begin{figure}
\centerline{\psfig{file=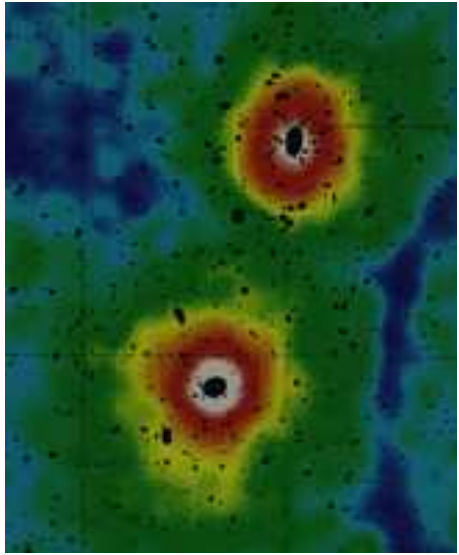,width=6cm,clip=}}
%\vskip 7.5cm
\caption{FIGURE 1. ROSAT/PSPC image of A3528 (Schindler 1996). The
superposed galaxy 
distribution shows all galaxies observed in direction of the cluster (not only cluster
galaxies). While in the optical one sees only an enhancement 
of the galaxy density, the X-ray
image shows clearly the shape of the cluster potential -- in this case
a bimodal structure.
}
\end{figure}
%---------------------------------

-- ROSAT observations and also earlier missions revealed 
   that clusters have very
   different morphologies: some have substructure (see
   Figs.~1, 2
   and 4) while others show a spherically symmetric X-ray emission
   (see Fig.~3). These are 
   indications for the different dynamical states: in the former case
   merging processes are taking place while the latter clusters are
   already in virial equilibrium. As different cosmological models
   predict a different percentage of virialized clusters, one can in
   principle determine $\Omega_0$ with morphological studies
   (Richstone et al. 1992).

-- From X-ray spectra, observed mainly with ASCA, one can determine
   the temperature of the cluster. The temperatures (2-10 keV) are
   generally in
   good agreement with the depth of the cluster potential
   well. Temperature variations over the cluster (see e.g. Markevitch
   et al. 1998) indicate the dynamical state of the cluster and can
   also be used to discriminate between different cosmological
   models. Another interesting parameter one can determine with X-ray
   spectra is the metallicity of the intra-cluster gas. Typical
   metallicities are between 0.2 and 0.5 in solar units (e.g. Mushotzky \&
   Loewenstein 1997; Tsuru et al. 1997), indicating
   that at least part of the gas cannot be primordial but must have
   been processed in the cluster galaxies. The distribution of the
   metals within the cluster and the ratios between different elements
   give therefore  
   important hints on the origin of the intra-cluster gas.

-- In several clusters the Sunyaev-Zel'dovich effect was detected
   (Birkinshaw 1998),
   i.e. photons of the cosmic microwave
   background are inverse Compton scattered when they pass through the hot
   intracluster gas so that the apparent brightness of the cosmic microwave
   background in direction of the cluster is changed (Sunyaev \&
   Zel'dovich 1972).  
   A combination of this effect with an X-ray observation
   yields an estimate for the physical
   size of the cluster. Comparing it with the angular size one gets a
   direct estimate of the distance of the cluster, completely
   independent from any other distance indicator.

-- Assuming spherical symmetry one can determine from X-ray
   observations the amount of gas in a
   cluster. With the additional assumption of hydrostatic equilibrium
   one can estimate as well the total mass of the cluster, which is a
   very interesting quantity, because about 70-90\% of the cluster
   mass is dark matter. The combination of both mass profiles 
   yields the radial
   distribution of the dark matter. The ratio of gas mass and total
   mass gives 
   $\Omega_{baryon}$, assuming that the matter is accumulated
   indiscriminately in the cluster potential wells. 
   The values found for $\Omega_{baryon}$ are so high that for
   an $\Omega=1$ universe they are 
   in contradiction with primordial nucleosynthesis, a
   problem named ``baryonic catastrophe'' (see e.g. White et al. 1993).

-- As clusters are the largest bound objects in the universe they trace
   the mass distribution of the universe very well and are thus excellent
   tracers for large-scale structure. Therefore statistical
   studies of the cluster distribution and cluster properties are very powerful
   to constrain cosmological parameters.

In the following I present as an example for a nearby cluster the
Virgo cluster (Sect. 2.1) and two examples for distant clusters --
RXJ1347.5-1145 and Cl0939+4713 (Sect. 2.2). In Sect. 3 X-ray
properties of several clusters are compared and some (preliminary)
cosmological conclusions are drawn.

\bsk
\ni 2. CLUSTERS AT DIFFERENT REDSHIFTS  
\ssk
\ni

\bsk
\ni
2.1 Morphology of the Virgo cluster: gas versus galaxies
\ssk
\ni

As the Virgo cluster is the nearest cluster it is very well studied --
both in X-rays and in optical. Therefore it is the ideal object
for a detailed comparison of the two components -- the gas and the
galaxies. We use the X-ray
emission from the ROSAT All-Sky Survey (B\"ohringer et al. 1994) and
the optical information 
from the Virgo Cluster Catalog (Binggeli et al. 1985). 

Fig.~2 shows a comparison of the X-ray emission of the gas and the
galaxy distribution. It is obvious that the
cluster is very irregular. There are several substructures, which are very
similar in both components:
the main X-ray maximum at M87 and a second maximum around M86 are
connected by a region with the highest galaxy density. In the south,
around M49, another maximum is visible -- both in X-ray and in the galaxy
density. South-west of M87 is a relatively steep slope visible in both
components. 

%--------------------------  figure 2
\begin{figure}
\centerline{\psfig{file=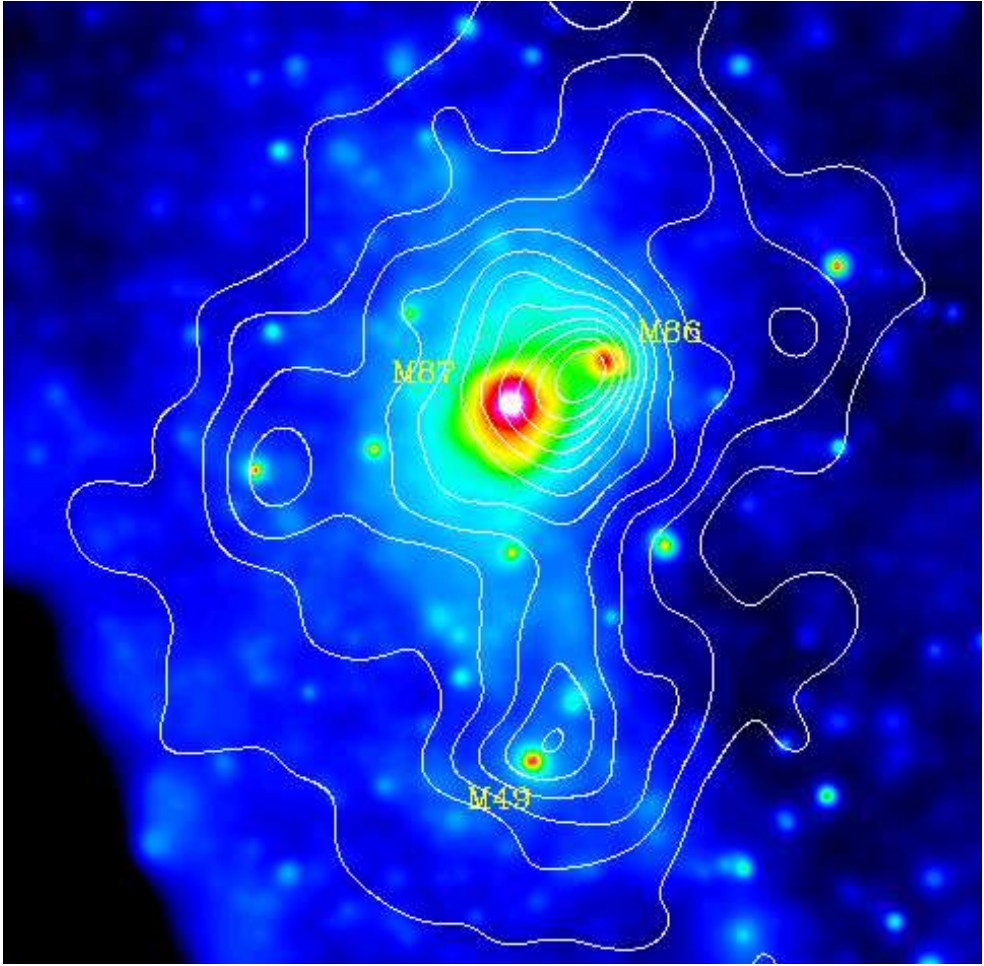,width=13.cm,clip=}}
\caption{FIGURE 2. Comparison of the X-ray and the optical appearance of the
Virgo cluster. The colour image shows
the X-ray emission as observed in the ROSAT All-Sky Survey. The
contours show the number density of the 1292 
member galaxies of the Virgo Cluster Catalog (Binggeli et al. 1995). 
The galaxy distribution is smoothed with a Gaussian of
$\sigma$ = 24'. The size of the image is $12^{\circ}\times 12^{\circ}$.
}
\end{figure}
%---------------------------------

For a more quantitative analysis we decomposed the three different
subclusters around M87, M49, and M86 and compared their
three-dimensional density profiles
individually. We found a large difference between the gas profile and
the profile of the galaxy distribution in all subclusters 
The gas profiles are 
much steeper in the centre (up to a radius of at least 80 arcminutes) 
and somewhat shallower in the outer
regions than the galaxy profile.

Subdividing the galaxy sample into different morphological types confirms the
different distributions for early and late-type galaxies found by
Binggeli et al. (1997). The spiral and irregular galaxies have a very
flat profile, while the elliptical galaxies have a steep profile,
indicating that they are more concentrated on the subcluster centres. 
But although their distribution is relatively steep compared to the
rest of the galaxies, it is still much flatter than the X-ray profile
in particular in the centre.

Comparing the profiles of the different subclusters one finds
a systematic effect: the smaller the subcluster, the steeper the
profile, both in X-rays and optical. This effect together with the
very similar appearance of the two components shows that some kind of
equilibrium state must have established within the potential wells of
the subclusters although the cluster has such a irregular structure. 
For details see Schindler et al. (1998b).

%--------------------------  figure 3
\begin{figure}
\centerline{\psfig{file=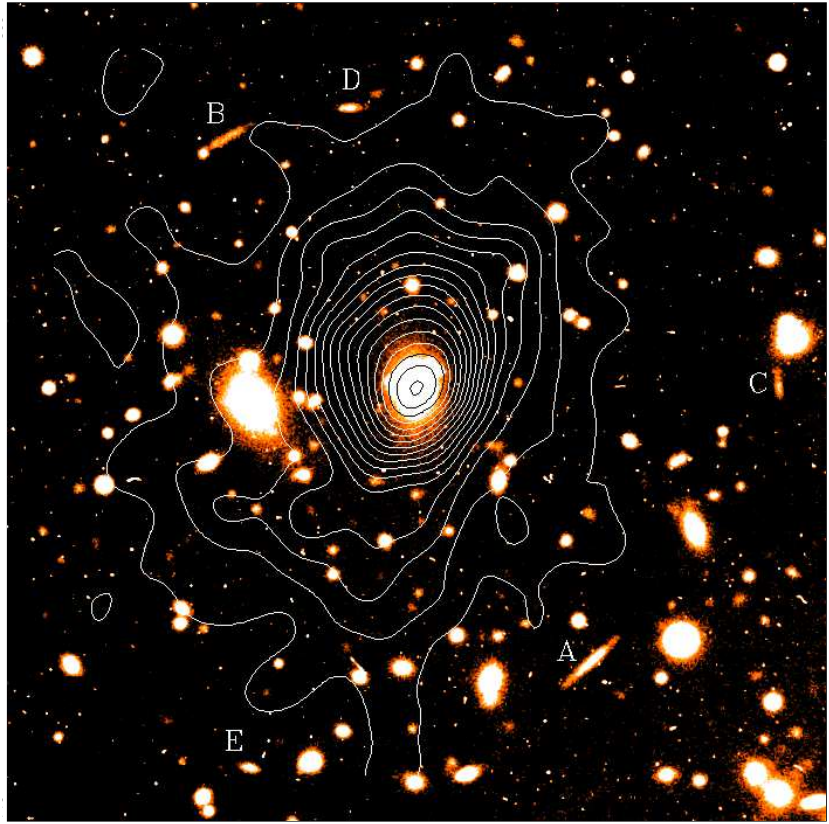,width=11.cm,clip=}}
\caption{FIGURE 3. ROSAT/HRI contours  of RXJ1347.5-1145
superposed on an optical image. The X-ray emission is strongly peaked.
In the optical image one can see that the cluster acts as a
gravitational lens.
The positions of the gravitational arcs are marked with letters. 
}
\end{figure}
%---------------------------------
\bsk
\ni
2.2 Properties of distant clusters
\ssk
\ni
Exemplarily, I will present two clusters at about the same redshift,
but with very different properties.

RXJ1347.5-1145 at z=0.45 
is an exceptional cluster in many respects (Schindler
et al. 1997). With a
luminosity of $L_{X,bol}=2.1\times10^{46}$erg/s it is the most
luminous X-ray cluster found so far. The X-ray image (Fig.~3) is strongly
peaked which suggests that there is a cooling flow in the centre of
the cluster. In a standard cooling flow analysis one finds a mass
accretion rate of more than 3000 ${\cal M}_\odot$ per year, which is
also an extreme value. 

The optical image (Fig.~3) taken with the NTT reveals several arcs,
indicating that the cluster is acting as a gravitational lens. The
large distance of the arcs from the cluster centre of 240 kpc shows
that the cluster is very massive. From ASCA observations we found a
relatively high gas temperature of 9.3 keV and a metallicity of 0.33
in solar units.

All these properties suggest that this cluster is dynamically old: 
well virialized structure, indications for high mass, early metal
enrichment, no merger in the recent past.

%--------------------------  figure 4
\begin{figure}
\centerline{\psfig{file=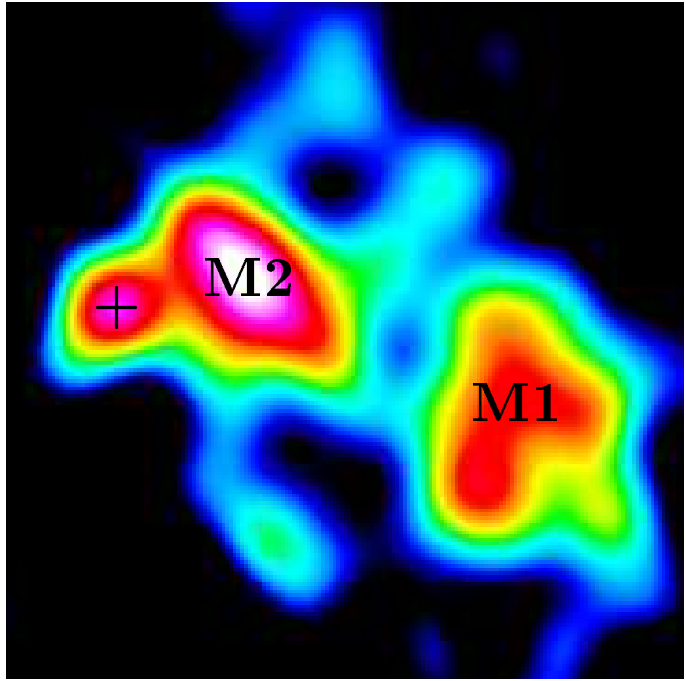,width=9.cm,clip=}}
\caption{FIGURE 4. ROSAT/HRI image of the Cl0939+4713. The cluster
consists of two subclusters (marked M1 and M2) which have even some
internal structure. The source marked with a cross is a background
quasar at a redshift of 2.
}
\end{figure}
%---------------------------------

The cluster Cl0939+4713  -- also known as the
most distant Abell cluster A851 --  
is at about the same redshift (z=0.41) and
shows also the gravitational lens effect, i.e. is also a massive cluster.
But already the ROSAT/HRI image (Fig.~4) shows that the cluster is
very different (Schindler et al. 1998a). 
It consists of two subclusters which have
even some internal structure. Obviously, this cluster is in a merging
process. The X-ray luminosity is more than an order of
magnitude lower than the one of RXJ1347.5-1145  
and the metallicity is even compatible with 0 (determined with an ASCA
observation). This is all the more
surprising as the cluster has so many galaxies.
While it is claimed to be the optically richest
cluster, RXJ1347.5-1145 is relatively poor in terms of galaxies.

From all its X-ray properties one would conclude that Cl0939+4713 is a
dynamically young cluster.

In Table 1 a few properties are summarized. The table is complemented
by a few other clusters which cannot be discussed explicitly because
of lack of space. But I would like to draw your attention on the
puzzling cluster AXJ2019
(Hattori et al. 1997) -- a distant cluster at a redshift of 1.0 which
shows an Fe line corresponding to a supersolar abundance.
\vskip 0.3cm

\begin{tabular}{|l|c|c|c|c|c|c|c|}
\noalign{\hrule height 0.8pt}
& & & & & & & \cr
& nearby &AC118&Cl0500& Cl0939& RXJ1347&Cl0016&AXJ2019$^a$\cr
& & & & & & & \cr
\noalign{\hrule height 0.8pt}
& & & & & & & \cr
redshift &     &0.31& 0.32& 0.41 &0.45& 0.55 & 1.0       \cr
& & & & & & & \cr
$L_{X,bol}$& 0.05-5&6.8&0.6& 1.6 & 21 &5.0$^b$& 1.9    \cr
& & & & & & & \cr
metallicity &0.2-0.5 &0.23$^c$& 0.0-1.5$^d$&0.22 &  0.33&0.11$^e$& $\approx
1.7$\cr
& & & & & & & \cr
temperature& 2-10   &9.3$^c$& 7.2$^d$& 7.6 & 9.3 & 8.0$^e$ & 8.6  \cr 
& & & & & & & \cr
substructure &in 25\%$^f$ & yes & yes& yes & no  & yes & ?    \cr
& & & & & & & \cr
\noalign{\hrule height 0.8pt}
\end{tabular}
\vskip 0.3cm
\caption{TABLE 1. Comparison of some properties of nearby and distant 
clusters. The clusters are
ordered according to their redshift. The properties in the second
column are average values for nearby clusters. The X-ray luminosity is
in units of $10^{45}$erg/s and the temperature is in keV.
$^a$ from Hattori et al. (1997), $^b$ from Neumann
\& B\"ohringer (1997), 
$^c$ from Mushotzky \& Loewenstein (1997), $^d$ from Ota et
al. (1998), $^e$ from Furuzawa et al. (1998), $^f$ from Neumann (1997)} 
}

\bsk
\ni 3. CONCLUSIONS
\ssk
\ni 

A comparison of several properties at different
redshifts (Table 1) shows in none of the properties a 
clear trend with time. Similar results
were found e.g. by Mushotzky \& Loewenstein (1997) and Tsuru et
al. (1997). These findings suggest an early evolution of galaxy
clusters, which is almost completed at a redshift of 0.5-1.0. This
results points a low $\Omega$ universe, because in a
high $\Omega$ universe one would expect a strong evolution in this
period. Of course this result is very preliminary given the small
sample of high-redshift clusters.

Another obvious feature in Table 1 is the strong variation in the
properties, e.g. in the X-ray luminosity which varies over more than
an order of magnitude. The optical richness does not seem to have any
correlation with the metallicity or the X-ray luminosity. 
In some properties like in the metallicity the large scatter might be
caused by the large uncertainties in the metallicity determination,
but in other parameters like the X-ray luminosity the scatter is certainly  
real. Obviously,
cluster evolution is quite complex, i.e. not every cluster is evolving
in the same way with the same initial ingredients. 
For understanding the cluster evolution we need
clearly larger samples of more distant clusters and properties like
the metallicity determined to a higher accuracy. We hope that all 
these goals will
be achieved by the next X-ray missions XMM, AXAF and ASTRO-E.

%\bsk
%\baselineskip = 12pt
%{\abstract \ni ACKNOWLEDGMENTS
%}

\bsk
\baselineskip = 12pt

% beginning of font "references"

{\references \ni REFERENCES
\ssk

\ref Birkinshaw M., Physics Reports, in press (astro-ph/9808050)
\ref Binggeli B., Sandage A., Tammann G.A., 1985, AJ 90, 1681
\ref Binggeli B., Tammann G.A., Sandage A., 1987, AJ 94, 251
\ref B\"ohringer H., Briel U.G., Schwarz R.A., Voges W., Hartner G.,
       Tr\"umper J., 1994, Nature 368, 828
\ref Furuzawa A., Tawara, Y., Kunieda, H., Yamashita, K.,
        Sonobe T., Tanaka, Y., Mushotzky R., 1998, ApJ 504, 35
\ref Hattori, M., Ikebe, Y., Asaoka, I., Takeshima, T., B\"ohringer, H.,
     Mihara, T., Neumann, D.M., Schindler, S., Tsuru, T., Tamura, T., 1997,
     Nature 388, 146
\ref Markevitch M., Forman W.R., Sarazin C.L., Vikhlinin A., 1998, ApJ
        503, 77
\ref Mushotzky R.F., Loewenstein M., 1997, ApJ 481, L63
\ref Neumann, D.M., B\"ohringer, H. 1997, MNRAS, 289, 123
\ref Neumann, D.M. 1997, Ph.D. Thesis,
        Ludwigs-Maximilians-Universit\"at, M\"unchen
\ref Ota N., Mitsuda K., Fukazawa Y., 1998, ApJ 495, 170
\ref Richstone D., Loeb A., Turner E.L., 1992, ApJ 393, 477
\ref Schindler S., 1996, MNRAS 280, 309
%\ref Schindler S., Wambsganss J., 1997, A\&A, 322, 66
\ref Schindler S., Hattori M., Neumann D.M., B\"ohringer H., 1997,
     A\&A, 317, 646
\ref Schindler S., Belloni P., Ikebe Y., Hattori M., Wambsganss J.,
       Tanaka Y., 1998a, A\&A, 338, 843
\ref Schindler S., Binggeli B., B\"ohringer H., 1998b, A\&A, submitted
\ref Sunyaev R.A., Zel'dovich Y.B., 1972, Comments Astrophys. Space
       Sci. 4, 173
\ref Tsuru T.G., Matsumoto H., Koyama K., Tomida H., Fukazawa Y., 
        Hattori M., Hughes J.P., 1997, astro-ph/9711353 
\ref  White S.D.M., Navarro J.F., Evrard A.E., Frenk C.S., 1993, Nature
      366, 429

}                      

\end{document}